\documentclass[12pt]{iopart}

\usepackage{bm}
\usepackage{mathrsfs}
\usepackage{verbatim}
\usepackage[all,cmtip]{xy}
\usepackage{amssymb}
\usepackage{graphics}
\usepackage{amsthm}
\usepackage{color}
\usepackage{amscd}
\usepackage{amssymb}
\usepackage{color}
\usepackage{dsfont}
\definecolor{darkred}  {rgb}{0.5,0,0}
\definecolor{darkblue} {rgb}{0,0,0.5}
\definecolor{darkgreen}{rgb}{0,0.5,0}
\usepackage{hyperref}
\hypersetup{
	pdftitle = {QRT Proposal},
	pdfauthor = {},
	colorlinks = true,
	urlcolor  = blue,         
	linkcolor = red,     
	citecolor = blue,    
	filecolor = darkred       
}

\def\ra{\rangle}
\def\la{\langle}
\def\bb{\mathbb}

\def\ra{\rangle}
\def\la{\langle}
\def\bb{\mathbb}

\newtheorem{theorem}{Theorem}

\newtheorem{pro}[theorem]{Proposition}

\newtheorem{definition}{Definition}
\theoremstyle{remark}
\newtheorem{remark}{Remark}

\newcommand{\beax}{\begin{eqnarray*}}
\newcommand{\eeax}{\end{eqnarray*}}

\def\be{\begin{eqnarray}}
\def\ee{\end{eqnarray}}
\newcommand{\bea}{\begin{eqnarray}}
\newcommand{\eea}{\end{eqnarray}}

\newcommand{\mH}{\mathcal{H}}

\newcommand{\mS}{\mathcal{S}}

\def\>{\rangle}
\def\<{\langle}

\begin{document}

\title[Monogamy of quantum discord]{Monogamy of quantum discord}

\author{Yu Guo}

\address{Institute of Quantum Information Science, School of Mathematics and Statistics, Shanxi Datong University, Datong, Shanxi 037009, China}
\ead{guoyu3@aliyun.com}

\author{Lizhong Huang}

\address{Institute of Quantum Information Science, School of Mathematics and Statistics, Shanxi Datong University, Datong, Shanxi 037009, China}

\author{Yang Zhang}
\address{Institute of Theoretical Physics, Shanxi Datong University, Datong {\rm 037009}, China}

\begin{abstract}
The original quantum discord (QD) is shown to be not monogamous except for the three-qubit states.
Recently, a complete monogamy relation for multiparty quantum system was established for entanglement in [Phys. Rev. A. {101}, 032301~(2020)], and in addition, a new multipartite generalization of QD was proposed in [Phys. Rev. Lett. {124}, 110401~(2020)]. 
In this work, we firstly define the complete monogamy for this multipartite quantum discord (MQD) and the global quantum discord (GQD).
MQD, with the same spirit as the complete monogamy of entanglement, is said to be completely monogamous (i) if it does not increase under coarsening of subsystems and (ii) if some given combination of subsystems reach the total amount of the correlation, then all other combination of subsystems that do not include all the given subsystems
do not contain such a correlation any more. Here, coarsening of subsystems means discarding or combining some subsystems up to
the given partition. Simultaneously, the complete monogamy of GQD is also defined with slight modification on the coarsening relation.
Consequently, we explore all the coarsening relations of MQD and show that it is completely monogamous with a modicum assumption. In addition, with the same spirit, we investigate all the coarsening relations for GQD and show that GQD is not completely monogamous.
That is, in the sense of the complete monogamy relation, MQD as a generalization of the original QD captures the nature of such a quantum correlation, and thus it is nicer than GQD as a generalization.

\vspace{2pc}
\end{abstract}


\maketitle

\section{Introduction}

Quantum discord, as the foremost one of the quantum correlations
beyond entanglement, has been extensively explored in the last two decades
due to its remarkable applications in quantum information protocols~\cite{ollivier2001quantum,henderson2001classical,Datta2008prl,DiVincenzo2004prl, Sarandy2009pra,Xu2010nc,Rulli2011pra,Giorgi2011prl,Modi2012rmp,Bera2017epp,Radhakrishman2020prl,Hu2018pr,Zhou2019pra}.
It is originally defined for bipartite system as the minimized difference
between the quantum mutual information with and without
a von Neumann projective measurement applied on one of
the subsystems~\cite{ollivier2001quantum,henderson2001classical}.
Consequently, several multipartite generalizations
have been proposed based on different scenarios~\cite{Rulli2011pra,Giorgi2011prl,
	Radhakrishman2020prl}.
In Ref.~\cite{Rulli2011pra}, by means of the multipartite mutual information,
the global quantum discord is defined as the minimal difference between 
the mutual information the pre- and post-state under local von Neumann measurements.
In Ref.~\cite{Giorgi2011prl}, the total quantum discord and the genuine quantum discord
were investigated based on the total information
and the total classical correlation.
Very recently, Radhakrishnan \textit{et al.}~\cite{Radhakrishman2020prl} gave another way of 
extending quantum discord to multipartite case up to a fixed conditional local von Neumann measurement.
The global quantum discord, total quantum discord and the genuine quantum discord are symmetric under exchange
the subsystems, while the multipartite quantum discord in Ref.~\cite{Radhakrishman2020prl}
is asymmetric.

An important issue closely related to a nonlocal correlation measure of composite quantum system (such as entanglement measure, quantum discord, and Einstein-Podolsky-Rosen steering, etc.) is to explore
the distribution of the correlation over many parties~\cite{Coffman,streltsov2012are,Dhar,GG,GG2019,G2020,Gzy2020,Eltschka2019quantum,Deng}.
In this context, there are two ways to describe the distribution: one is the monogamy or polygamy relation by the bipartite measure~\cite{Coffman,streltsov2012are,Dhar,GG,GG2019} and the other one is the relation based on the multipartite measure~\cite{G2020,Gzy2020,Braga2012pra}.
Among these attempts, numerous efforts have been made for entanglement. However,
there are very few studies on quantum discord.   
For the bipartite measure of quantum discord, it is proved that the square of quantum discord obeys the monogamy relation for three qubit states~\cite{Ren2013qic} while other systems always display polygamous behavior
\cite{Ren2013qic,Prabhu2pra,Giorgi2011pra,Fanchini2013pra,Bai2013pra}.
In Ref.~\cite{Braga2012pra}, it has been shown that global quantum discord acts as a monogamy bound for pairwise quantum discord in which one of the subsystem is fixed while the other part runs over all subsystems
provided that bipartite discord does not increase under discard of subsystems.
According to the framework of complete monogamy relation (for entanglement),
when we deal with the multipartite systems, the traditional approach of exploring monogamy relation based the bipartite measure is not substantial. There are something missing~\cite{G2020}.
Therefore, for multipartite quantum discord and global quantum discord, we need to examine renewedly with the complete monogamy
protocol.

The aim of this article is to explore
the distribution of discord contained in a multipartite state whenever it is measured by multipartite quantum discord and global quantum discord respectively
in the framework of the complete monogamy relation.
However, quantum discord is far different from entanglement: quantum discord is not monotonic under local operation and classical communication (LOCC) while it is a basic feature for entanglement.
So we need at first to consider the coarsening relations when we consider monogamy of multipartite quantum discord and global quantum discord
and establish the framework of complete monogamy relation for multipartite quantum discord and global quantum discord.

The rest of this paper is organized as follows.
We recall the definitions of the original quantum discord, multipartite quantum discord, global quantum discord, and investigate coarser relation of multiparty partitions by which we then introduce the monogamy laws in
literatures in Section II. In Section III, we put forward the definition of complete multiparty quantum discord and by which we discuss and establish the framework of complete monogamy
relation for multipartite quantum discord and global quantum discord.
Section IV explores the coarsening relation and complete monogamy relation of
the multipartite quantum discord. It is divided into three subsections.
The first subsection discusses the tripartite case, the second subsection deals with
the four-partite case and the main conclusioin is given in the last subsection
for the general multipartite case. In Section V, we deal with global quantum discord in the framework of complete monogamy relation.
Finally, we conclude.
For convenience, throughout this paper,
we call the $n$-partite discord in Ref.~\cite{Radhakrishman2020prl}
multipartite quantum discord (MQD), and the one in Ref.~\cite{Rulli2011pra,Braga2012pra} global quantum discord (GQD), 
and we call both MQD and GQD multiparty quantum discord in order to differentiate these different concepts.

\section{Preliminaries}

In this section, we review at first the definitions of MQD and GQD, and then recall
the monogamy relation and complete monogamy relation in literatures.
Throughout this paper, we let $\mH^{A_1A_2\cdots A_n}$
be the Hilbert space corresponding to the $n$-parite quantum system with finite dimension.
and let $\mS^{X}$ be the set of density
operators acting on $\mH^{X}$.
The tripartite and the four-partite systems are always denoted by
$\mH^{ABC}$ and $\mH^{ABCD}$, respectively.

\subsection{Quantum discord and its generalization}

For any bipartite state $\rho\in\mS^{AB}$,
the original quantum discord is defined as~\cite{ollivier2001quantum,henderson2001classical}
\bea
D_{A;B}(\rho)=\min_{\Pi^A}\left[S_{B|\Pi^A}(\rho)-S_{B|A}(\rho)\right]
\label{eq:def_discord}
\eea
where $S_{B|A}(\rho)=S_{AB}(\rho)-S_{A}(\rho)$, $S(\rho)=-{\rm Tr}\rho\log\rho$ is the von Neumann entropy,
\beax
S_{B|\Pi^A} (\rho) = \sum_{j} p_{j}^AS_{AB}\left(\Pi_{j}^{A}\rho\Pi_{j}^{A}/p_{j}^A\right)=S_{AB}(\rho_{\Pi^A})- S_A(\rho_{\Pi^A}),
\eeax
$\Pi_{j}^{A}$ is a one-dimensional von Neumann projection operator on subsystem $ A $ and
$p_j^A={\rm Tr}(\Pi_{j}^A\rho\Pi_{j}^A)$,
$\rho_{\Pi^{A}}=\sum_{j}\Pi_{j}^{A}\rho\Pi_{j}^{A}$.
Hereafter, we denote $S(\rho^{X})$ by $S_{X}(\rho)$ sometimes for simplicity.

For multipartite systems, the $(n-1)$-partite measurement is written~\cite{Radhakrishman2020prl}
\bea
\Pi^{A_1\dots A_{n-1}}_{j_1 \dots j_{n-1}} = \Pi^{A_1}_{j_1} \otimes \Pi^{A_2}_{j_2|j_1}  \dots \otimes \Pi^{A_{n-1}}_{j_{n-1}|j_1\dots j_{n-2}} ,
\label{conditionalmeasurement}
\eea
where the $n$ subsystems are labeled as $A_i$, the measurements take place in the order $A_1 \rightarrow A_2 \rightarrow \dots A_{n-1}$.
For $\rho\in\mS^{A_1A_2\cdots A_n}$, the $n$-partite MQD is defined by~\cite{Radhakrishman2020prl}
\bea D_{A_1;A_2;  \dots ;A_n} (\rho)  &=&  \min_{ \Pi^{A_1 \dots A_{n-1} }} \Big[
	- S_{A_2 \dots A_n|A_1 }(\rho)
	+  S_{A_2| \Pi^{A_1} } (\rho)  \dots \nonumber\\
	&&~~~+ S_{A_n|  \Pi^{A_1 \dots A_{n-1} } }  (\rho) \Big]
	\label{multipartitediscord}
\eea
up to the measurement ordering $ A_1 \rightarrow A_2 \rightarrow \dots \rightarrow A_{n-1} $,
where
\beax S_{A_k|\Pi^{A_1 \dots A_{k-1}}} (\rho) =
\sum_{j_1 \dots j_{k-1}} p_{\bm{j}}^{(k-1)}
S_{A_1 \dots A_k}\left( \Pi^{(k-1)}_{\bm{j}}  \rho  \Pi^{(k-1)}_{\bm{j}}  / p_{\bm{j}}^{(k-1)} \right)
\eeax
with $\Pi^{(k)}_{\bm{j}}\equiv\Pi^{A_1 \dots A_{k}}_{j_1 \dots j_{k}}$,
$p_{\bm{j}}^{(k)}={\rm Tr}(\Pi^{(k)}_{\bm{j}}\rho\Pi^{(k)}_{\bm{j}})$.
For example, the triparite case $\rho\in\mS^{ABC}$, 
\beax
D_{A;B; C} (\rho)  &=&  \min_{ \Pi^{AB }} \Big[
- S_{BC|A}(\rho) +  S_{B| \Pi^{A} } (\rho)  + S_{C|  \Pi^{AB} }  (\rho) \Big]\\
&=&\min_{ \Pi^{AB }} \Big[
S(\rho'^{AB})-S(\rho'^{A}) +S(\rho'')-S(\rho''^{AB})-S(\rho)+S(\rho^{A})\Big]
\eeax
up to the measurement ordering $ A \rightarrow B$,
where $\Pi^A(\rho)=\rho'$, $\Pi^{AB}(\rho)=\rho''$, and $\rho^X={\rm Tr}_{\bar{X}}(\rho)$ with $\bar{X}$ denotes the complementary of $X$ in $\{A,B,C\}$ (e.g, $\rho^{AB}={\rm Tr}_C(\rho)$).

The GQD, denoted by ${D}_{A_1 : \cdots :A_n }$, for
an arbitrary state ${\rho}\in\mS^{A_1 \cdots A_n}$ under a set of local
measurements $\{{\Pi}^{A_1}_{j_1} \otimes \cdots \otimes {\Pi}^{A_n}_{j_n}\}$
is defined as~\cite{Rulli2011pra,Braga2012pra}
\begin{equation}
	{D}_{A_1 : \cdots : A_n}(\rho) = \min_{\Phi} \left[
	I(\rho) - I(\Phi(\rho)) \right],
\end{equation}
where
\begin{equation}\label{measurement-GQD}
	\Phi( {\rho}) = \sum_{k}\, {{\Pi}}_{k} {\rho} \,
	{{\Pi}}_{k},
\end{equation}
with ${\Pi}_{k} = {\Pi}^{A_1}_{j_1} \otimes \cdots \otimes {\Pi}^{A_n}_{j_n}$ and
$k$ denoting the index string $(j_1 \cdots j_n$), the mutual information $I({\rho})$
is defined by~\cite{Groisman2005pra}
\begin{eqnarray}
	I({\rho}) = \sum_{k=1}^{n} S_{A_k}\left( {\rho}\right) -
	S_{A_1 \cdots A_n}\left( {\rho} \right) ,
\end{eqnarray}
where
\begin{equation}
	\Phi \left( {\rho}^{A_k} \right) = \sum_{k^\prime} {\Pi}^{A_k}_{k^\prime} \, {\rho}^{A_k} \,
	{\Pi}^{A_k}_{k^\prime}.
\end{equation}
Note here that, the difference between $D_{A_1;A_2;  \dots ;A_n}$ and ${D}_{A_1 : \cdots :A_n }$ is that the former is semicolon in the subscript $A_1;A_2;  \dots ;A_n$
while the latter one is with colon in the subscript $A_1 : \cdots :A_n$.

\subsection{Coarser relation of multipartite partition}\label{coarser relation}

For any partition $X_1|X_2| \cdots |X_{k}$ of $A_1A_2\cdots A_n$ with $X_s=A_{s(1)}A_{s(2)}\cdots A_{s(f(s))}$, $s(i)<s(j)$ whenever $i<j$, and $s(p)<t(q)$ whenever $s<t$ for any possible $p$ and $q$, $1\leq s,t\leq k$.
For instance, partition $AB|C|DE$ is a $3$-partition of $ABCDE$.
Let $X_1|X_2| \cdots |X_{k}$ and $Y_1|Y_2| \cdots |Y_{l}$ be two partitions of $A_1A_2\cdots A_n$ or subsystem of $A_1A_2\cdots A_n$.
$Y_1|Y_2| \cdots |Y_{l}$ is said to be \textit{coarser} than $X_1|X_2| \cdots |X_{k}$, denoted by 
\bea
X_1|X_2| \cdots| X_{k}\succ Y_1|Y_2| \cdots |Y_{l}, 
\eea
if $Y_1|Y_2| \cdots |Y_{l}$ can be obtained from $X_1|X_2| \cdots| X_{k}$
by some or all of the following ways: 
\begin{itemize}
	\item (C1) Discarding some subsystem(s) of $X_1|X_2| \cdots| X_{k}$;
	\item (C2) Combining some subsystems of $X_1|X_2| \cdots| X_{k}$;
	\item (C3) Discarding some subsystem(s) of the last subsystem $X_k$ provided that the last subsystem $X_k$ in $X_1|X_2| \cdots| X_{k}$ is $X_{k}=A_{k(1)}A_{k(2)}\cdots A_{k(f(k))}$ with $f(k)\geq2$.
\end{itemize}
For example, $A|B|C|D|E \succ A|B|C|DE\succ A|B|C|D\succ AB|C|D\succ AB|CD$, $A|B|C|DE\succ A|B|DE$.
Clearly, $X_1|X_2| \cdots| X_{k}\succ Y_1|Y_2| \cdots |Y_{l}$ and $ Y_1|Y_2| \cdots |Y_{l}\succ Z_1|Z_2| \cdots| Z_{s}$ imply $X_1|X_2| \cdots| X_{k}\succ Z_1|Z_2| \cdots| Z_{s}$.
For any partition $X_1|X_2| \cdots| X_{k}$ of $A_1A_2\cdots A_n$ or subsystem of $A_1A_2\cdots A_n$,
the assciated MQD is $D_{X_1;X_2; \cdots; X_{k}}$ up to the measurement $\Pi^{X_1 \cdots X_{k-1}}$.
For example, for partition $A|B|CD$ of $ABCD$, the associated MQD is $D_{A;B;CD}(\rho^{ABCD})$ up to measurement
$\Pi^{A|B}$, while for partition $AB|C$ of the subsystem $ABC$ in $ABCD$, the associated MQD is $D_{AB;C}(\rho^{ABC})$ up to measurement
$\Pi^{AB}$, where $\Pi^{AB}$ is a von Neumann measurement acting on $AB$ with $AB$ is regarded as a sigle particle, $\rho^{ABC}={\rm Tr}_D(\rho^{ABCD})$. 
One can easily see that $X_1|X_2| \cdots| X_{k}\succ Y_1|Y_2| \cdots |Y_{l}$ if and only if
$\Pi^{Y_1 \dots Y_{l-1}}$ can be induced by $\Pi^{X_1 \dots X_{k-1}}$ but not vice versa.
In such a case, we say $\Pi^{Y_1 \dots Y_{l-1}}$ and $\Pi^{X_1 \dots X_{k-1}}$ are compatible.
Otherwsie, it is uncompatible.
For example, $\Pi^{A|BC}$ and $\Pi^{A|B}$ are uncompatible on the system $\mH^{ABCD}$
since the projection on part $BC$ (regared as a single part) is not ncessarily a projection on $B|C$.

Furthermore, if $X_1|X_2| \cdots| X_{k}\succ Y_1|Y_2| \cdots |Y_{l}$,
we denote by $\Xi(X_1|X_2| \cdots| X_{k}- Y_1|Y_2| \cdots |Y_{l})$ the set of
all the partitions that are coarser than $X_1|X_2| \cdots| X_{k}$ and either exclude any subsystem of $Y_1|Y_2| \cdots |Y_{l}$ or include some but not all subsystems of $Y_1|Y_2| \cdots |Y_{l}$.
We take the five-partite system $ABCDE$ for example,
\beax 
\Xi(A|B|CD|E-A|B)&=&\left\lbrace CD|E, A|CD|E,B|CD|E,A|CD, A|E,\right. \\
&&\left. B|E,A|C,A|D,B|C,B|D\right\rbrace ,
\eeax 
\beax
\Xi(A|B|C|D|E-A|C)&=&\left\lbrace B|D|E, B|D, B|E, D|E,A|B,A|D,A|E,\right. \\ &&B|C,C|D,C|E,A|B|D,A|BD,AB|D,A|B|E,\\
&&A|BE,AB|E,A|D|E,A|DE,AD|E,B|C|D,\\
&&B|CD,BC|D,B|C|E,B|CE,BC|E,B|D|E,\\
&&B|DE,BD|E,C|D|E,C|DE,CD|E,A|B|D|E,\\
&&B|C|D|E,AB|DE,BC|DE,A|BDE,ABD|E,\\
&&B|CDE,BCD|E,A|B|DE,AB|D|E,A|BD|E,\\
&&\left. BC|D|E,B|CD|E,B|C|DE\right\rbrace.
\eeax

For more clarity, we fix the following notations. Let $X_1|X_2| \cdots| X_{k}$ and $Y_1|Y_2| \cdots |Y_{l}$
be partitions of $A_1A_2\cdots A_n$ or subsystem of $A_1A_2\cdots A_n$.
We denote by 
\bea
X_1|X_2| \cdots| X_{k}\succ^a Y_1|Y_2| \cdots |Y_{l}
\eea 
for the case of (C1) 
and by,
\bea
X_1|X_2| \cdots| X_{k}\succ^b Y_1|Y_2| \cdots |Y_{l}
\eea 
for the case of of (C2), and in addition by
\bea
X_1|X_2| \cdots| X_{k}\succ^c Y_1|Y_2| \cdots |Y_{l}
\eea 
for the case of (C3). For example, $A|B|C\succ^a A|B$, $A|B|C\succ^b A|BC$, $A|BC\succ^c A|B$, $A|BC\succ^c A|C$.

\subsection{Complete multipartite entanglement measure}

The counterpart to MQD for multipartite entanglement is the complete multipartite entanglement measure.
A function 
$E^{(n)}: \mS^{A_1A_2\cdots A_n}\to\bb{R}_{+}$ 
is called an $n$-partite entanglement measure in 
literature~\cite{Horodecki2009} if
it satisfies:
\begin{itemize}
	\item {\bf(E1)} $E^{(n)}(\rho)=0$ if $\rho$ is fully separable;
	\item {\bf(E2)}
	$E^{(n)}$ cannot increase
	under $n$-partite LOCC.
\end{itemize}
When we take into consideration an $n$-partite measure of entanglement or other quantum correlation, we need 
discuss whether it is defined uniformly for any 
$k$-partite system at first, $k<n$. 
Let $Q^{(n)}$ be a multipartite measure (for entanglement or quantum discord, ect). If $Q^{(k)}$ is uniquely determined by $Q^{(n)}$ for any $2\leq k<n$, then we call $Q^{(n)}$ a \textit{uniform} measure. 
For example, MQD and GQD are uniquely defined for any $k$, thus they are uniform measures. The $n$-partite entanglement of formation~\cite{G2020} defined as
$E_f^{(n)}( |\psi\rangle)=\frac12\sum_{i=1}^mS( \rho^{A_i}) $
for pure state and via the convex-roof extension for mixed states (i.e.,
$E_{f}^{(n)}(\rho):=\min\sum_{i} p_iE_f^{(n)}(|\psi_i\ra)$
for any mixed state $\rho$, where the minimum is taken over all pure-state
decomposition $\{p_i,|\psi_i\ra\}$ of $\rho$),
is a uniform multipartite entanglement measure.
That is, a uniform measure is series of measures that have uniform expressions definitely.
A uniform multipartite entanglement measure $E^{(n)}$ is called a \textit{unified}
multipartite entanglement measure if it also satisfies the following
condition~\cite{G2020}: 
\begin{itemize}
	\item {\bf(E3)}~\textit{the unification condition}, i.e., 
	$E^{(n)}$ is consistent with $E^{(k)}$ for any $2\leqslant k<n$.
\end{itemize}
The unification condition should be comprehended in the following sense~\cite{G2020}.
Let 
$|\psi\ra^{A_1A_2\cdots A_n}=|\psi\ra^{A_1A_2\cdots A_k}|\psi\ra^{A_{k+1}\cdots A_n}$, then
\beax
&&E^{(n)}(|\psi\ra^{A_1A_2\cdots A_n})\\
&=&E^{(k)}(|\psi\ra^{A_1A_2\cdots A_k})+E^{(n-k)}|\psi\ra^{A_{k+1}\cdots A_n}.
\eeax
And
\beax
E^{(n)}(\rho^{A_1A_2\cdots A_n})=E^{(n)}(\rho^{\pi(A_1A_2\cdots A_n)})
\eeax 
for any $\rho^{A_1A_2\cdots A_n}\in\mS^{A_1A_2\cdots A_n}$, where $\pi$ is a permutation of the subsystems. 
In addition, 
\beax
E^{(k)}(X_1|X_2| \cdots| X_{k})\geqslant E^{(l)}(Y_1|Y_2| \cdots |Y_{l})
\eeax
for any $\rho^{A_1A_2\cdots A_n}\in\mS^{A_1A_2\cdots A_n}$ whenever $X_1|X_2| \cdots| X_{k}\succ^a Y_1|Y_2| \cdots |Y_{l}$,
where the vertical bar indicates the split
across which the entanglement is measured.
A uniform MEM $E^{(n)}$ is called a \textit{complete}
multipartite entanglement measure if it satisfies both {\bf(E3)} above and the following~\cite{G2020}: 
\begin{itemize}
	\item {\bf(E4)}~$E^{(n)}(X_1|X_2| \cdots| X_{k})\geqslant
	E^{(k)}(Y_1|Y_2| \cdots |Y_{l})$
	holds for all $\rho\in\mS^{A_1A_2\cdots A_n}$ whenever $X_1|X_2| \cdots| X_{k}\succ^b Y_1|Y_2| \cdots |Y_{l}$.
\end{itemize}
It is easy to see that $E_f^{(n)}$ is a complete multipartite entanglement measure~\cite{G2020}.

\subsection{Monogamy relation}

It is known that
classical correlations can be freely shared among many parties, i.e., A
party A can have maximal classical correlations with
two parties B and C simultaneously. But this is no longer the
case if quantum entanglement or other nonlocal correlations are
concerned~\cite{Dhar}. The impossibility of sharing those
types of nonclassical correlations unconditionally across many parties are known as monogamy
constraints.
For a given bipartite measure of nonlocal correlation $Q$, $Q$ is said to be monogamous (we take the tripartite case for example) if~\cite{Coffman,Dhar}
\bea\label{monogamy1}
Q(A|BC)\geq Q(AB)+Q(AC).
\eea
However, Eq.~(\ref{monogamy1}) is not valid for many entanglement measures~\cite{Coffman,Dhar} and quantum discord~\cite{Ren2013qic} but some power function
of $Q$ admits the monogamy relation, i.e., $Q^\alpha(A|BC)\geq Q^\alpha(AB)+Q^\alpha(AC)$ for some $\alpha>0$.
In Ref.~\cite{GG}, we address this issue by proposing an improved definition of monogamy (without inequalities) for entanglement measure:
A measure of entanglement $E$ is monogamous if for any $\rho\in\mS^{ABC}$
that satisfies the \textit{disentangling condition}, i.e.,
\be\label{cond}
E( \rho^{A|BC}) =E( \rho^{AB}),
\ee
we have that $E(\rho^{AC})=0$. With respect to this definition, a continuous measure $E$ is monogamous according to this definition if and only if
there exists $0<\alpha<\infty$ such that
\be\label{power}
E^\alpha( \rho^{A|BC}) \geq E^\alpha( \rho^{AB}) +E^\alpha( \rho^{AC}),	
\ee
for all $\rho$ acting on the state space
$\mH^{ABC}$
with fixed $\dim\mH^{ABC}=d<\infty$ (see Theorem 1 in Ref.~\cite{GG}).
With this improved definition of monogamy, we proved that almost all the bipartite entanglement measures so far are monogamous~\cite{GG,GG2019}.
Notice that, for
these bipartite measures, only the
relation between $A|BC$, $AB$ and $AC$ are revealed, the global correlation
in ABC and the correlation contained in part $BC$ is missed~\cite{G2020}. That is, the monogamy relation in such a sense is not ``complete''. Recently, we established a complete monogamy relation for entanglement in Ref.~\cite{G2020}.
For a unified tripartite entanglement measure $E^{(3)}$,
it is said to be {completely monogamous} if is satisfies the complete disentangling condition~\cite{G2020}, i.e.,
for any
$\rho\in\mathcal{S}^{ABC}$ that satisfies\be\label{condofm2}
E^{(3)}(\rho^{ABC})=E^{(2)}(\rho^{AB})
\ee
we have that $E^{(2)}(\rho^{AC})=E^{(2)}(\rho^{BC})=0$.
If $E^{(3)}$ is a continuous unified tripartite
entanglement measure. Then, $E^{(3)}$ is completely monogamous
if and only if there exists
$0<\alpha<\infty$ such that~\cite{G2020}
\bea\label{power2}
E^{\alpha}(\rho^{ABC})\geq  E^{\alpha}(\rho^{AB})
+ E^{\alpha}(\rho^{AC})+ E^{\alpha}(\rho^{BC}),
\eea
for all $\rho^{ABC}\in\mathcal{S}^{ABC}$ with fixed $\dim\mH^{ABC}=d<\infty$,
here we omitted the superscript $^{(2,3)}$ of $E^{(2,3)}$ for brevity.

Let $E^{(3)}$ be a complete MEM. $E^{(3)}$ is defined to be
tightly complete monogamous if it satisfies the tight disentangling conditon, i.e.,
for any state $\rho^{ABC}\in\mS^{ABC}$ that satisfying~\cite{G2020}
\bea\label{condofm5}
E^{(3)}(\rho^{ABC})=E^{(2)}(\rho^{A|BC})
\eea
we have $E^{(2)}(\rho^{BC})=0$,
which is equivalent to
\beax
E^{\alpha}(\rho^{ABC})\geqslant E^{\alpha}(\rho^{A|BC})+E^{\alpha}(\rho^{BC})
\eeax
for some $\alpha>0$, here we omitted the superscript $^{(2,3)}$ of $E^{(2,3)}$ for brevity.
For the general case of $E^{(m)}$, one can similarly process with the same spirit.

\section{Framework of complete monogamy relation for quantum discord}

All the previous discussions on the monogamy for quantum discord in literatures are based on the bipartite quantum discord as in Eq.~(\ref{monogamy1}).
The complete monogamy relation in Eq.~(\ref{condofm2}) displays the distribution of the correlation throughly.
We thus adopt this scenario to describe the monogamy of MQD and GQD, namely, we consider the complete monogamy relation as Eqs.~(\ref{condofm2}) and ~(\ref{condofm5})
with MQD/GQD replacing entanglement measure $E$.

According to the complete multipartite entanglement measure, for a uniform measure of quantum correlation $Q^{(m)}$, we expect it satisfies the 
following conditions when we discuss the complete monogamy relation [we take the tripartite case for example, the $m$-partite ($m\geq4$) case can be argued similarly]:
\bea
(\textbf{U1}): Q^{(3)}(\rho^{AB}\otimes \rho^{C})=Q^{(2)}(\rho^{AB}),\quad \forall~\rho^{AB}\otimes\rho^C\in\mS^{ABC}.
\eea
\bea
(\textbf{U2}): Q^{(3)}(\rho^{ABC})=Q^{(3)}(\rho^{\pi(ABC)}),\quad \forall~\rho^{ABC}\in\mS^{ABC}.
\eea  
\bea
(\textbf{U3}): Q^{(3)}(ABC)\geq Q^{(2)}(XY),\quad \forall~\rho^{ABC}\in\mS^{ABC}.
\eea
\bea
(\textbf{U4}): Q^{(3)}(ABC)\geq Q^{(2)}(X|YZ),\quad \forall~\rho^{ABC}\in\mS^{ABC}.
\eea

We now begin to check whether MQD and GQD are unified.
It is proved in Ref.~\cite{Radhakrishman2020prl} that $D_{A;B;C}=D_{X;Y}$ for any $\rho=\rho^{XY}\otimes\rho^Z\in\mS^{ABC}$, $\{X,Y,Z\}=\{A,B,C\}$. Namely, MQD satisfies (\textbf{U1}). Going further,
for any bipartite state $\rho\in\mS^{AB}$, for any given decomposition
$\rho=\sum_ip_i\rho_i^{AB}$, it can be extended by adding an auxiliary system $C$ that does not correlate
with $AB$ as
\bea
\rho=\sum_ip_i|i\ra\la i|^C\otimes \rho_i^{AB}.
\eea
That is, $D_{C;AB}=0$.
One can easily show that
\bea\label{unification1}
D_{C;A;B}=D_{A;B}
\eea
for such a state.
We can also show that (see in Proposition~\ref{tripartite-trade-off} and Proposition~\ref{tripartite-complete-monogamy} in the next Section)
\bea\label{unification2}
D_{A;C;B}\geq D_{A;B}
\eea
with the equality holds if and only if $D_{A;C}=0$
for such a state.
Eqs.~(\ref{unification1},\ref{unification2})
imply that: (i) when an auxiliary particle classically correlated with the state is added and it does not disturb the measurement ordering $A\rightarrow B$, then the MQD equals to the pre-state;
(ii) when the auxiliary particle disturb the previous measurement ordering $A\rightarrow B$, then the multipartite discord does not decrease.
That is
MQD obeys more than (\textbf{U1}).
But MQD violates (\textbf{U2}) obviously.
The GQD satisfies (\textbf{U1}) since
$D_{A:B:C}(\rho^{AB}\otimes\rho^C)=D_{A:B}(\rho^{AB})$ and (\textbf{U2}) is clear since it is symmetric.
However, for these measurement-induced correlations, whether they obey
conditions (\textbf{U3}) and (\textbf{U4}),
is not straightforward.

We are now ready to discuss the complete monogamy of these two generalizations of quantum discord.
With the same principle as the unified/complete multiparty entanglement measure and the complete monogamy of entanglement introduced in the previous section, we can now give the definition
of complete multiparty quantum discord and complete monogamy for MQD and GQD.

\begin{definition}\label{De-cqd}
	Let $D_{A_1|A_2|\cdots|A_n}$ be $D_{A_1;A_2;\cdots;A_n}$ or $D_{A_1:A_2:\cdots:A_n}$ defined on $\mS^{A_1A_2\cdots A_n}$. With the notations aforementioned, $D_{A_1|A_2|\cdots|A_n}$ is said to be complete if
	it is monotonic under coarsening of subsystem(s), i.e., for any $\rho\in\mS^{A_1A_2\cdots A_n}$,
	we have 
		\bea\label{cqd}
		D_{X_1|X_2|\cdots|X_{k}}\geq D_{Y_1|Y_2|\cdots|Y_{l}}
		\eea
		holds for any partitions $X_1|X_2| \cdots |X_{k}$ and $Y_1|Y_2| \cdots |Y_{l}$ of $A_1A_2\cdots A_m$ or subsystem of $A_1A_2\cdots A_m$ with $X_1|X_2| \cdots| X_{k}\succ Y_1|Y_2| \cdots |Y_{l}$ whenever  $D_{A_1|A_2|\cdots|A_n}=D_{A_1;A_2;\cdots;A_n}$ and with $X_1|X_2| \cdots| X_{k}\succ^{a,b} Y_1|Y_2| \cdots |Y_{l}$ whenever $D_{A_1|A_2|\cdots|A_n}=D_{A_1:A_2:\cdots:A_n}$. 
\end{definition}

\begin{definition}\label{De-monogamy}
	Let $D_{A_1|A_2|\cdots|A_n}$ be $D_{A_1;A_2;\cdots;A_n}$ or $D_{A_1:A_2:\cdots:A_n}$ defined on $\mS^{A_1A_2\cdots A_n}$. With the notations aforementioned, $D_{A_1|A_2|\cdots|A_n}$ is said to be completely monogamous if it is complete and 
	satisfies the dis-correlated condition, i.e., for any state $\rho\in\mS^{A_1A_2\cdots A_n}$ that satisfies 
		\bea\label{dis-correlated}
		{D}_{X_1|X_2| \cdots|X_{k}}={D}_{Y_1|Y_2|\cdots|Y_{l}}
		\eea
		we have that
		\bea
		D_{\Gamma}=0
		\eea
		holds for all $\Gamma\in \Xi(X_1|X_2| \cdots| X_{k}- Y_1|Y_2| \cdots |Y_{l})$, where $X_1|X_2| \cdots |X_{k}$ and $Y_1|Y_2| \cdots |Y_{l}$ are arbitrarily given partitions of $A_1A_2\cdots A_m$ or subsystem of $A_1A_2\cdots A_m$, and where $X_1|X_2| \cdots| X_{k}\succ Y_1|Y_2| \cdots |Y_{l}$ whenever $D_{A_1|A_2|\cdots|A_n}=D_{A_1;A_2;\cdots;A_n}$, $X_1|X_2| \cdots| X_{k}\succ^{a,b} Y_1|Y_2| \cdots |Y_{l}$ whenever $D_{A_1|A_2|\cdots|A_n}=D_{A_1:A_2:\cdots:A_n}$.
\end{definition}

That is, MQD/GQD is complete if and only if it satisfies the conditions (\textbf{U3}) and (\textbf{U4}). 
Namely, for MQD and GQD, we do not distinguish conditions (\textbf{U3}) and (\textbf{U4}) which are corresponding to the unified measure and complete measure for entanglement respectively, and call it complete for both of them uniformly. 
In other words, MQD/GQD is a complete multiparty quantum discord if Eq.~(\ref{cqd}) hold true.
Condition~(\ref{dis-correlated}) is the counterpart to the complete disentangling condition together with the tight disentangling condition for multipartite entanglement.
It is not necessary to distinguish these two disentangling conditions for
MQD and GQD. In what follows, both the complete disentangling condition and the tight disentangling condition for multipartite entanglement are called complete disentangling condition for simplicity.

We illustrate Definitions~\ref{De-cqd} and~\ref{De-monogamy} with the four partite case.
$D_{A;B;C;D}$ is complete if: For any state $\rho\in\mS^{ABCD}$, ${D}_{A;B;C;D}\geq{D}_{X;Y;Z}\geq{D}_{M;N}$ for any $\{M,N\}\subseteq\{X,Y,Z\}\subseteq\{A,B,C,D\}$ with ordering $A\rightarrow B\rightarrow C\rightarrow D$ for both $M, N$ and $X, Y, Z$. In addition, ${D}_{A;B;C;D}\geq{D}_{A;BC;D}\geq{D}_{A;BC}\geq D_{A;B}$,
	${D}_{A;B;C;D}\geq{D}_{A;BCD}\geq{D}_{A;BC}\geq D_{A;C}$, 	${D}_{A;B;C;D}\geq{D}_{AB;CD}\geq{D}_{AB;C}$, ${D}_{AB;CD}\geq{D}_{AB;D}$,
	${D}_{A;B;C;D}\geq{D}_{ABC;D}$, ${D}_{A;B;C;D}\geq{D}_{AB;C;D}\geq{D}_{AB;C}$, ${D}_{AB;C;D}\geq{D}_{AB;D}$, ${D}_{A;B;C;D}\geq{D}_{A;B;CD}\geq{D}_{AB;CD}$,
	${D}_{A;B;CD}\geq{D}_{A;B;C}$, ${D}_{A;B;CD}\geq{D}_{A;B;D}$.
$D_{A;B;C;D}$ is completely monogamous if: For any state $\rho\in\mS^{ABCD}$ that satisfies ${D}_{A;B;C;D}={D}_{A;B;C}$, we have that
	$D_{A;B;D}=D_{A;C;D}=D_{B;C;D}=D_{A;D}=D_{B;D}=D_{C;D}=0$. For any state $\rho\in\mS^{ABCD}$ that satisfies ${D}_{A;B;C;D}={D}_{A;B}$, we have that
	$D_{C;D}=D_{A;C;D}=D_{B;C;D}=D_{A;C}=D_{A;D}=D_{B;C}=D_{B;D}=0$.
	Similarly, for any state $\rho\in\mS^{ABCD}$ that satisfies ${D}_{A;B;C;D}={D}_{A;C}$ , we have that
	$D_{A;B}=D_{B;D}=D_{A;B;D}=D_{B;C;D}=D_{A;D}=D_{B;C}=D_{C;D}=0$.
	In addition, for any state $\rho\in\mS^{ABCD}$ that satisfies ${D}_{A;B;C;D}={D}_{AB;C}$ , we have that
	$D_{A;B}=D_{AB;D}=D_{C;D}=0$.
	The other cases can be argued similarly.
$D_{A:B:C:D}$ is complete if:
For any state $\rho\in\mS^{ABCD}$, ${D}_{A:B:C:D}\geq{D}_{X:Y:Z}\geq{D}_{M:N}$ for any $\{M,N\}\subseteq\{X,Y,Z\}\subseteq\{A,B,C,D\}$.
	In addition, ${D}_{A:B:C:D}\geq{D}_{A:BC:D}\geq{D}_{A:BC}$,
	${D}_{A:B:C:D}\geq{D}_{A:BCD}$, 	${D}_{A:B:C:D}\geq{D}_{AB:CD}$, 
	${D}_{A:B:C:D}\geq{D}_{ABC:D}$, ${D}_{A:B:C:D}\geq{D}_{AB:C:D}\geq{D}_{AB:C}$, ${D}_{AB:C:D}\geq{D}_{AB:D}$, ${D}_{A:B:C:D}\geq{D}_{A:B:CD}\geq{D}_{AB:CD}$.
$D_{A:B:C:D}$ is completely monogamous if: For any state $\rho\in\mS^{ABCD}$ that satisfies ${D}_{A:B:C:D}={D}_{A:B:C}$, we have that
	$D_{A:B:D}=D_{A:C:D}=D_{B:C:D}=D_{A:D}=D_{B:D}=D_{C:D}=0$. For any state $\rho\in\mS^{ABCD}$ that satisfies ${D}_{A:B:C:D}={D}_{A:B}$, we have that
	$D_{C:D}=D_{A:C:D}=D_{B:C:D}=D_{A:C}=D_{A:D}=D_{B:C}=D_{B:D}=0$.
	Similarly, for any state $\rho\in\mS^{ABCD}$ that satisfies ${D}_{A:B:C:D}={D}_{A:C}$ , we have that
	$D_{A:B}=D_{B:D}=D_{A:B:D}=D_{B:C:D}=D_{A:D}=D_{B:C}=D_{C:D}=0$.
	In addition, for any state $\rho\in\mS^{ABCD}$ that satisfies ${D}_{A:B:C:D}={D}_{AB:C}$ , we have that
	$D_{A:B}=D_{AB:D}=D_{C:D}=0$.
	The other cases can be processed analogously.

\begin{remark}
	We call it complete monogamy here for MQD and GQD since both MQD and GQD are multipartite measures, and this is consistent with
	the complete and tightly complete monogamy for multipartite entanglement measure.
	The previous research in this field is based on the bipartite measure of quantum discord
	\cite{Ren2013qic,Prabhu2pra,Giorgi2011pra,Fanchini2013pra,Bai2013pra} and it is called monogamy but not complete monogamy.
	This is analogous to that of bipartite entanglement measure in which it is called monogamy but not complete monogamy.
	That is, for both quantum discord and entanglement, `complete' refers to the uniform multipartite measure and the one without `complete' refers to the bipartite measure.
\end{remark}

\begin{remark}
	Comparing with the definition of the complete monogamy for entanglement, the difference is that, for entanglement we only require the condition~(\ref{dis-correlated}), i.e., the complete disentangling condition for entanglement (see Eqs.~(\ref{cond}) and (\ref{condofm2})).
	Condition~(\ref{cqd}) is true naturally for any unified multipartite entanglement measure $E^{(m)}$ since it is decreasing under tracing out any subsystem. But this fact is not obvious for other quantum correlation. So the monogamy of 
	quantum discord need this assumption necessarily. 
\end{remark}

\begin{remark}
	For MQD, Eq.~(\ref{cqd}) in Definition~\ref{De-cqd} and Eq.~(\ref{dis-correlated}) in Definition~\ref{De-monogamy} are 
	presented under the fixed ording $ A_1 \rightarrow A_2 \rightarrow \dots \rightarrow A_{n-1}\rightarrow A_{n}$ since MQD strictly depends on the ordering of the measurements on subsystems.
	But it is superfluous for the globle quantum discord since it admits $(\textbf{U2})$.
\end{remark}

\begin{remark}
	In Definition~\ref{De-cqd} and Definition~\ref{De-monogamy}, MQD and GQD are different: The coarser relation of class (C3) is meaningless for GQD since in such a case the corresponding measurements are not compatible as we argued in Subsection~\ref{coarser relation}.
	It has been shown that there exists three-qubit state $\rho\in\mS^{ABC}$ such that $D_{A:BC}<D_{A:B}$~\cite{Braga2012pra}. 
	However, althouhgh $D_{A:BC}<D_{A:B}$ for some state, it seems true that $D_{A:B:C}>D_{M:N}$ for any tripartite state (also see the assumption in Proposition~\ref{GQD-1}), $\{M,N\}\subseteq \{A,B,C\}$.
	That is, the monotonicity should holds whenever the measurement on both sides are the same one. But for the case of MQD, such a case can not occur since the last subsystem is not measured, so the remaining measurements are always compatible. This is why we require three classes of coarsening relations when we camparing $D_{X_1;X_2; \cdots; X_{k}}$ and $D_{Y_1;Y_2; \cdots; Y_{l}}$, and only require coarsening classes of (C1) and (C2) for that of GQD in these definitions. 	   
\end{remark}

\section{Monogamy of the multipartite quantum discord}\label{4}

For convenience, we fix some notations.
For any local measurement $\Pi^{A_{i_1}\cdots A_{i_k}}$ acting on the reduced state $\rho^{A_{i_1}\cdots A_{i_k}A_{i_{k+1}}}$ of $\rho^{A_1A_2\cdots A_{n}}\in\mS^{A_1A_2\cdots A_{n}}$,
the conditional mutual information changes by an amount
\be
\begin{array}{rcl}
	d_{A_{i_1}\dots A_{i_k};A_{i_{k+1}}}=S_{A_{i_{k+1}}|\Pi^{A_{i_1}\dots A_{i_k}}}-S_{A_{i_{k+1}}|A_{i_1}\dots A_{i_k}},
\end{array}
\ee
where $1\leq i_1<i_2<\cdots i_k<i_{k+1}\leq n$.

\subsection{The tripartite case}

Different from multipartite entanglement, it is unknown whether the correlation decreases
under coarsening the subsystems.

\begin{pro}\label{tripartite-trade-off}
	For any state $\rho$ in $\mS^{ABC}$, the following coarsening relations hold: 
	\begin{itemize}
		\item $D_{A;B;C}\geq D_{A;B}+D_{AB;C}$, $D_{A;B;C}\geq D_{A;C}$.
		\item $D_{A;B;C} \geq D_{B;C}$ provided that $d_{AB;C}\geq d_{B;C}$.
		\item $D_{A;B;C}\geq D_{A;BC}\geq D_{A;B}$.
		\item $D_{A;BC}\geq D_{A;C}$ provided that $d_{AB;C}\geq d_{A;C}$.
	\end{itemize}
That is, if $d_{AB;C}\geq d_{B;C}$ and $d_{AB;C}\geq d_{A;C}$, then $D_{A;B;C}$ is a complete multiparty quantum discord.
\end{pro}

\begin{proof}
	We assume that $D_{A;B} (\rho^{AB})=S(\rho'^{AB})-S(\rho'^{A})-S(\rho^{AB})+S(\rho^{A})$
	for some $\Pi^A(\rho^{AB})=\rho'^{AB}$,
	and that
	$D_{A;B;C} (\rho^{ABC})=S(\rho''^{AB})-S(\rho''^{A})+S(\rho'''^{ABC})-S(\rho'''^{AB})-S(\rho^{ABC})+S(\rho^{A})$ for some $\Pi'^{AB}$ with $\Pi'^A(\rho^{ABC})=\rho''^{ABC}$ and $\Pi'^{AB}(\rho^{ABC})=\rho'''^{ABC}$. (Hereafter, in this subsection, we denote by $\rho'$ the state after the optimal local measurement on one particle of the bipartite state, by $\rho''$ the state after the optimal local measurement on one particle of the tripartite state, and by $\rho'''$ the state after the optimal local measurement on two particles of the bipartite state.)
	For simplicity, we denote $S(\rho^{*X})$ by $S^*_X$ hereafter (e.g.,
	we denote $S(\rho''^{AB})$ by $S''_{AB}$ for brevity). 
	It is straightforward that, for any given $\rho\in\mS^{ABC}$, we have
	\beax
	&&D_{A;B;C}  -D_{A;B} \\
	&=&   \min_{\Pi^{AB}  } (  - S_{BC|A}
	+ S_{B| \Pi^{A}}
	+ S_{C|  \Pi^{AB}}  )
	-\min_{ \Pi^A } (S_{B|\Pi^A}  - S_{B|A}  )\\
	&=&(S''_{AB}-S''_{A}+S'''_{ABC}-S'''_{AB}
	-S_{ABC}+S_{A})
	-(S''_{AB}-S''_{A}-S_{AB}+S_{A})\\
	&&+(S''_{AB}-S''_{A}-S_{AB}+S_{A})
	-(S'_{AB}-S'_{A}-S_{AB}+S_{A})\\
	&\geq&(S''_{AB}-S''_{A}+S'''_{ABC}-S'''_{AB}
	-S_{ABC}+S_{A})
	-(S''_{AB}-S''_{A}-S_{AB}+S_{A})\\
	&=&S'''_{ABC}-S'''_{AB}-S_{ABC}+S_{AB}\\
	&\geq&D_{AB;C}.
	\eeax
	Note that, even thought the optimal local measurement does not exist, we have that, for any $\epsilon>0$,
	there exist $\Pi^A$ and $\Pi'^{AB}$ such that
	$D_{A;B;C}\geq S(\rho''^{AB})-S(\rho''^{A})+S(\rho'''^{ABC})-S(\rho'''^{AB})-S(\rho^{ABC})+S(\rho^{A})-\epsilon$
	and $S(\rho''^{AB})-S(\rho''^{A})-S(\rho^{AB})+S(\rho^{A})\geq D_{A;B}$.
	It turns out that $D_{A;B;C}-D_{A;B}\geq\left[ S(\rho''^{AB})-S(\rho''^{A})+S(\rho'''^{ABC})-S(\rho'''^{AB})-S(\rho^{ABC})+S(\rho^{A})\right] -\left[S(\rho''^{AB})-S(\rho''^{A})-S(\rho^{AB})+S(\rho^{A}) \right] -\epsilon\geq0$ is equivalent to $[ S(\rho''^{AB})-S(\rho''^{A})+S(\rho'''^{ABC})-S(\rho'''^{AB})-S(\rho^{ABC})+S(\rho^{A})] -\left[S(\rho''^{AB})-S(\rho''^{A})-S(\rho^{AB})+S(\rho^{A}) \right]\geq0$ since $\epsilon$ can be arbitrarily small. So we can assume with no loss of generality that the optimal local measurement always exists hereafter.
	Similarly,
	\beax
	&&D_{A;B;C}  -D_{A;C} \\
	&\geq&(S''_{AB}-S''_{A}+S'''_{ABC}-S'''_{AB}
	-S_{ABC}+S_{A})
	-(S''_{AC}-S''_{A}-S_{AC}+S_{A})\\
	&=&(S'''_{ABC}-S'''_{AB}-S_{ABC}+S_{AB})
	+(S''_{ABC}-S''_{AC}-S_{ABC}+S_{AC})\\
	&&-(S''_{ABC}-S''_{AB}-S_{ABC}+S_{AB})\\
	&\geq&(S'''_{ABC}-S'''_{AB}-S_{ABC}+S_{AB})
	-(S''_{ABC}-S''_{AB}-S_{ABC}+S_{AB})\\
	&=&S'''_{ABC}-S'''_{AB}-S''_{ABC}+S''_{AB}\\
	&\geq&0.
	\eeax
	Here, the second inequality holds since the mutual information always decreases under local operation~\cite{Nielsen} (which is equivalent to $S''_{XY}-S''_{X}-S_{XY}+S_{X}\geq0$).

	Observe that
	\beax\label{3-partite-trade-off}
	&&D_{A;B;C}  -D_{B;C}\nonumber\\
	&\geq&(S''_{AB}-S''_{A}+S'''_{ABC}-S'''_{AB}
	-S_{ABC}+S_{A})
	-(S'''_{BC}-S'''_{B}-S''_{BC}+S''_{B})\nonumber\\
	&=&(S''_{AB}-S''_{A}+S'''_{ABC}-S'''_{AB}
	-S_{ABC}+S_{A})
	-(S'''_{BC}-S'''_{B}-S_{BC}+S_{B})\nonumber\\
	&=&(S'''_{ABC}-S'''_{AB}-S_{ABC}+S_{AB})
	+(S''_{AB}-S''_{A}-S_{AB}+S_{A})\nonumber\\
	&&-(S'''_{BC}-S'''_{B}-S_{BC}+S_{B}),
	\eeax
	thus $D_{A;B;C} \geq D_{B;C} $ since
	$d_{AB;C}\geq d_{B;C}$ by assumption.
	The following calculation are clear.
	$D_{A;B;C}  -D_{A;BC}\geq(S''_{AB}-S''_{A}+S'''_{ABC}-S'''_{AB}
	-S_{ABC}+S_{A})
	-(S''_{ABC}-S''_{A}-S_{ABC}+S_{A})=(S''_{AB}-S''_{A}+S'''_{ABC}-S'''_{AB}
	-S_{ABC}+S_{A})
	-(S''_{ABC}-S''_{A}-S_{ABC}+S_{A})
	=(S'''_{ABC}-S'''_{AB}-S''_{ABC}+S''_{AB})\geq 0$
	implies $D_{A;B;C} \geq D_{A;BC}$.
	$D_{A;BC}  -D_{A;B}
	\geq(S''_{ABC}-S''_{A}-S_{ABC}+S_{A})
	-(S''_{AB}-S''_{A}-S_{AB}+S_{A})
	=(S''_{ABC}-S''_{AB}-S_{ABC}+S_{AB})
	\geq0$, that is, $D_{A;BC} \geq D_{A;B}$.
	$D_{A;BC}  -D_{A;C}
	\geq(S''_{ABC}-S''_{A}-S_{ABC}+S_{A})
	-(S''_{AB}-S''_{A}-S_{AC}+S_{A})
	\geq0$ since
	$d_{AB;C}\geq d_{A;C}$ by assumption.
	This completes the proof.	
\end{proof}

\begin{pro}\label{tripartite-complete-monogamy}
	$D_{A;B;C}$ is completely monogamous if it is complete.
\end{pro}

\begin{proof}	
	We use the notations in the proof of Proposition~\ref{tripartite-trade-off}.
	
	Case 1: $D_{A;B;C}=D_{A;B}$. If $D_{A;B;C} (\rho^{ABC})=D_{A;B}(\rho^{AB})$,  we get $S(\rho'''^{ABC})-S(\rho'''^{AB})-S(\rho^{ABC})+S(\rho^{AB})=0$ for some von Neumann measurement $\Pi^{AB}(\rho^{ABC})=\rho'''^{ABC}$, which implies that
	$\rho^{ABC}=\sum_{k,j}p_{k,j}|k\ra\la k|^{A}\otimes|j\ra\la j|^B\otimes \rho_{k,j}^C$.
	It is clear that $D_{A;C}=D_{B;C}=0$ for such a state.
	
	Case 2: $D_{A;B;C}=D_{B;C}$.
	According to Eq.~(\ref{3-partite-trade-off}), for any state $\rho\in\mS^{ABC}$,
	if $D_{A;B;C}=D_{B;C}$, then
	$S''_{ABC}-S''_{A}-S_{ABC}+S_{A}=I_{BC|A}-I''_{BC|A}=0$.
	Therefore
	$\rho=\sum_{k}p_{k}|k\ra\la k|^{A}\otimes\rho_{k}^{BC}$ for some
	basis $\{|k\ra^A\}$ of $\mH^A$, and thus
	$D_{A;B}=D_{A;C}=0$.
	
	Case 3: If $D_{A;B;C}=D_{A;C}$, one can easily check that
	$\rho=\sum_{k,j}p_{k,j}|k\ra\la k|^{A}\otimes|j\ra\la j|^B\otimes \rho_{k,j}^C$
	for some basis $\{|k\ra^A |j\ra^B\}$ in $\mH^{AB}$.
	It follows that $D_{A;B}=D_{B;C}=0$.
	
	The other two cases $D_{A;B;C}=D_{A;BC}$ and $D_{A;B;C}=D_{A;BC}$ can be checked analogously.
\end{proof}

Using the same notations as in the proof of Proposition~\ref{tripartite-trade-off},	it is clear that,
for any tripartite state $\rho\in\mS^{ABC}$,
\beax
&&D_{A;B;C}  -D_{A;B} -D_{A;C}\\
&\geq&(S'''_{ABC}-S'''_{AB}-S_{ABC}+S_{AB})
-(S'''^{AC}-S'''^{A}-S_{AC}+S_{A})\\
&\geq&0
\eeax
whenever $d_{AB;C}\geq d_{A;C}$. That is

\begin{pro}
	For any state $\rho\in\mS^{ABC}$ that satisfies $d_{AB;C}\geq d_{A;C}$, we have
	\bea
	D_{A;B;C}\geq D_{A;B}+D_{A;C}.
	\eea	
\end{pro}

\subsection{The four-partite case}

With the increasing of particles involved, the hierarchy of coarsening relations become more complicated.

\begin{pro}\label{4partite-trade-off}
	For any state $\rho$ in $\mS^{ABCD}$, the following coarsening relations hold: \begin{itemize}
		\item $D_{A;B;C;D}\geq D_{A;B;C}+D_{ABC;D}$.
		\item $D_{A;B;C;D}\geq D_{A;B;D}$.
		\item $D_{A;B;C;D}\geq D_{A;B}$, $D_{A;B;C;D}\geq D_{A;C}$, $D_{A;B;C;D} \geq D_{A;D} $.
		\item $D_{A;B;C;D}\geq D_{A;C;D}$ provided that $d_{ABC;D}\geq d_{AC;D}$.
		\item $D_{A;B;C;D}\geq D_{A;C;D}+D_{A;B}$ provided that $d_{ABC;D}\geq d_{AC;D}$ and $d_{AB;C}\geq d_{A;C}$.
		\item $D_{A;B;C;D}\geq D_{B;C;D}+D_{A;B}$ provided that $d_{ABC;D}\geq d_{BC;D}$ and $d_{AB;C}\geq d_{B;C}$.
		\item $D_{A;B;C;D}\geq D_{A;B;D}+D_{AB;C}$ provided that $d_{ABC;D}\geq d_{AB;D}$.
		\item $D_{A;B;C;D} \geq D_{B;C} $ provided that $d_{AB;C}\geq d_{B;C}$ or $d_{ABC;D}\geq d_{B;C}$.
		\item $D_{A;B;C;D} \geq D_{B;D} $ provided that $d_{ABC;D}\geq d_{B;D}$.
		\item $D_{A;B;C;D} \geq D_{C;D} $ provided that $d_{ABC;D}\geq d_{A;D}$.
		\item $D_{A;B;C;D} \geq D_{AB;CD}+D_{A;B}$.
		\item $D_{A;B;C;D} \geq D_{A;BCD}$.
		\item $D_{A;B;C;D} \geq D_{A;BC;D}$.
		\item $D_{A;B;C;D} \geq D_{AB;C;D}+D_{A;B}$.
		\item $D_{A;B;C;D} \geq D_{A;B;CD}$.
	\end{itemize}
That is, $D_{A;B;C;D}$ is a complete multiparty quantum discord if $d_{ABC;D}\geq d_{AB;D}$, $d_{ABC;D}\geq d_{AC;D}$, 
$d_{ABC;D}\geq d_{BC;D}$, $d_{AB;C}\geq d_{A;C}$, and $d_{AB;C}\geq d_{B;C}$. 
\end{pro}

\begin{proof}
	For any $\rho\in\mS^{ABCD}$,
	we assume that 
	$D_{A;B;C;D} (\rho^{ABCD})=S(\rho'^{AB})-S(\rho'^{A})+S(\rho''^{ABC})-S(\rho''^{AB})+S(\rho'''^{ABCD})-S(\rho''^{ABC})-S(\rho^{ABCD})+S(\rho^{A})$ for some $\Pi^{ABC}$ with $\Pi^A(\rho^{ABCD})=\rho'^{ABCD}$, $\Pi^{AB}(\rho^{ABCD})=\rho''^{ABCD}$ and $\Pi^{ABC}(\rho^{ABCD})=\rho'''^{ABCD}$. (Hereafter, in this subsection, we denote by $\rho'$ the state after the optimal local measurement on one particle of the four-partite state, by $\rho''$ the state after the optimal local measurement on two particles of the four-partite state, and by $\rho'''$ the state after the optimal local measurement on three particles of the four-partite state.)
	Obviously, $D_{A;B;C}(\rho^{ABC})\leq S'_{AB}-S'_{A}+S''_{ABC}-S''_{AB}-S_{BC}+S^{A}$ and $S_{ABCD}'''-S_{ABC}''' -S_{ABCD}+S_{ABC}\geq D_{ABC;D}$.
	It is straightforward that
	\beax
	&&D_{A;B;C:D}  -D_{A;B;C} \\
	&\geq&   (  -S_{ABCD}+S_A+S_{AB}'-S_A'
	+S_{ABC}''-S_{AB}''+S_{ABCD}'''-S_{ABC}'''
	)\\
	&&- (S'_{AB}-S'_{A}+S''_{ABC}
	-S''_{AB}
	-S_{ABC}+S^{A})\\
	&=&(S_{ABCD}'''-S_{ABC}''' -S_{ABCD}+S_{ABC} )
	+(S_{ABC}''-S_{AB}'' -S_{ABC}+S_{AB} )\\
	&&+(S_{AB}'-S_{A}'-S_{AB} +S_{A} )
	-(S_{ABC}''-S_{AB}'' -S_{ABC}+S_{AB})\\
	&&-(S_{AB}'-S_{A}'-S_{AB} +S_{A})\\
	&=&S_{ABCD}'''-S_{ABC}''' -S_{ABCD}+S_{ABC} \\
	&\geq&D_{ABC;D}
	\eeax
	and
	\beax
	&&D_{A;B;C:D}  -D_{A;B;D} \\
	&\geq&(S_{ABCD}'''-S_{ABC}''' -S_{ABCD}+S_{ABC} )
	+(S_{ABC}''-S_{AB}'' -S_{ABC}+S_{AB} )\\
	&&+(S_{AB}'-S_{A}'-S_{AB} +S_{A} )-(S_{ABD}''-S_{AB}'' -S_{ABD}+S_{AB})\\
	&&-(S_{AB}'-S_{A}'-S_{AB} +S_{A})\\
	&=&(S_{ABCD}'''-S_{ABC}''' -S_{ABCD}+S_{ABC} )
	-(S_{ABD}''-S_{AB}'' -S_{ABD}+S_{AB})\\
	&&+(S_{ABC}''-S_{AB}'' -S_{ABC}+S_{AB} )\\
	&=&(S_{ABCD}'''-S_{ABC}''' -S_{ABCD}''+S_{ABC}'' )\\
	&&+(S_{ABCD}''-S_{ABD}'' -S_{ABCD}+S_{ABD} )\\
	&\geq &0.
	\eeax
	Similarly,
	\beax
	&&D_{A;B;C:D}  -D_{A;C;D} \\
	&\geq&(S_{ABCD}'''-S_{ABC}''' -S_{ABCD}+S_{ABC} )
	+(S_{ABC}''-S_{AB}'' -S_{ABC}+S_{AB} )\\
	&&+(S_{AB}'-S_{A}'-S_{AB} +S_{A} )-(S_{ACD}''-S_{AC}'' -S_{ACD}+S_{AC})\\
	&&-(S_{AC}'-S_{A}'-S_{AC} +S_{A})\\
	&=&(S_{ABCD}'''-S_{ABC}''' -S_{ABCD}+S_{ABC} )
	-(S_{ACD}''-S_{AC}'' -S_{ACD}+S_{AC})\\
	&&+(S_{ABC}''-S_{AB}''-S_{ABC}+S_{AB} )
	+(S_{AB}'-S_{A}'-S_{AB} +S_{A} )\\
	&&-(S_{AC}'-S_{A}'-S_{AC} +S_{A})\\
	&=& (S_{ABCD}'''-S_{ABC}''' -S_{ABCD}+S_{ABC} )
	-(S_{ACD}''-S_{AC}'' -S_{ACD}+S_{AC})\\
	&&+(S_{ABC}''-S_{AB}''-S_{ABC}'+S_{AB}' )+(S_{ABC}'-S_{AC}'-S_{ABC}+S_{AC} )\\
	&\geq&(S_{ABCD}'''-S_{ABC}''' -S_{ABCD}+S_{ABC} )
	-(S_{ACD}''-S_{AC}'' -S_{ACD}+S_{AC})\\
	&\geq&0
	\eeax
	provided that $d_{ABC;D}\geq d_{AC;D}$.
	Evidently, $D_{A;B;C:D}  -D_{A;C;D}\geq D_{A;B}$ whenever $d_{ABC;D}\geq d_{AC;D}$ and $d_{AB;C}\geq d_{A;C}$.
	With the same argument, one can check that $D_{A;B;C;D}\geq D_{B;C;D}+D_{A;B}$ provided that $d_{ABC;D}\geq d_{BC;D}$ and $d_{AB;C}\geq d_{B;C}$.
	Since $D_{A;B;C;D}\geq D_{A;B;C}$ and $D_{A;B;C}\geq D_{A;B}$, $D_{A;B;C}\geq D_{A;C}$,
	we get $D_{A;B;C;D}\geq D_{A;B}$ and $D_{A;B;C;D}\geq D_{A;C}$ directly.
	We also have $D_{A;B;C:D}\geq D_{A;D}$ since
	\beax
	&&D_{A;B;C:D}  -D_{A;D} \\
	&\geq&(S_{ABCD}'''-S_{ABC}''' -S_{ABCD}+S_{ABC} )
	+(S_{ABC}''-S_{AB}'' -S_{ABC}+S_{AB} )\\
	&&+(S_{AB}'-S_{A}'-S_{AB} +S_{A} )-(S_{AD}'-S_{A}'-S_{AD}+S_{A})\\
	&=&(S_{ABCD}'''-S_{ABC}''' -S_{ABCD}'+S_{ABC}' )
	+(S_{ABCD}'-S_{AD}' -S_{ABCD}+S_{AD})\\
	&&+(S_{ABC}''-S_{AB}''-S_{ABC}'+S_{AB}' )\\
	&\geq &0.
	\eeax

	The other items can also be easily checked, and thus the proof is completed.
\end{proof}

\begin{pro}\label{monogamy-4}
	$D_{A;B;C;D}$ is completely monogamous if it is complete, i.e.,
  \begin{itemize}
  	\item $d_{ABC;D}\geq d_{XY;D}$ for $\{X,Y\}\in\{\{A,C\}, \{B,C\}\}$.
  	\item $d_{ABC;D}\geq d_{Z;D}$ for $Z\in\{A, B,\}$.
  	\item $d_{AB;C}\geq d_{B;C}$
  \end{itemize}
	hold for any state $\rho\in\mS^{ABCD}$.	
\end{pro}

\begin{proof}
	For convenience, we use the same notations as that of the proof of Proposition~\ref{4partite-trade-off}.
	If $D_{A;B;C;D} (\rho^{ABCD}) =D_{A;B} (\rho^{AB})$,
	then $S_{ABCD}'''-S_{ABC}'''-S_{ABCD}+S_{ABC} =0$ and
	$S_{ABC}'-S_{AB}' -S_{ABC}+S_{AB} =0$, which implies that
	$\rho^{ABCD}=\sum_{k,j,l}p_{k,j,l}|k\ra\la k|^{A}\otimes|j\ra\la j|^B\otimes|l\ra\la l|^C\otimes \rho_{k,j,l}^D$.
	Thus $D_{A;C}=D_{A;D}=D_{B;C}=D_{B;D}=D_{C;D}=0$.
	If $D_{A;B;C;D} (\rho^{ABCD}) =D_{A;B;C} (\rho^{ABC})$,
	then $S_{ABCD}'''-S_{ABC}''' -S_{ABCD}+S_{ABC} =0$, which implies that
	$\rho^{ABCD}=\sum_{k,j,l}p_{k,j,l}|k\ra\la k|^{A}\otimes|j\ra\la j|^B\otimes|l\ra\la l|^C\otimes \rho_{k,j,l}^D$.
	Thus $D_{A;D}=D_{B;D}=D_{C;D}=0$.
	Similarly, one can easily verify that
	all the other dis-correlate conditions are valid.
\end{proof}

\subsection{The $n$-partite case}

Moving to general $n$-partite system, we can conclude the following theorems which are main results of this paper by similar arguments as that of Propositions~\ref{4partite-trade-off} and~\ref{monogamy-4}.

\begin{theorem}\label{th1}
	Let $D_{A_1;A_2;\cdots;A_n}$ be the $n$-partite quantum discord defined on $\mS^{A_1A_2\cdots A_n}$. Then the following holds true for any state $\rho\in\mS^{A_1A_2\cdots A_n}$:
	\begin{itemize}
		\item $D_{A_1;A_2;\cdots;A_n}\geq D_{A_1A_2\cdots A_{n-1};A_n}+D_{A_1;A_2;\cdots;A_{n-1}}$.
		\item $D_{A_1;A_2;\cdots;A_n}\geq D_{A_1;A_2;\cdots; A_p;A_3}$, $2\leq p\leq n-1$.
		\item $D_{A_1;A_2;\cdots;A_n}\geq D_
		{A_1;A_i}$, $3\leq i\leq n$.
		\item $D_{X_1;X_2; \cdots; X_{k}}\geq D_{Y_1;Y_2; \cdots; Y_{l}}$ for any $X_1|X_2| \cdots| X_{k}\succ^b Y_1|Y_2| \cdots |Y_{l}$. Moreover, $D_{X_1;X_2; \cdots; X_{k}}\geq D_{Y_1;Y_2; \cdots; Y_{l}}+D_{X_1;X_2; \cdots; X_{q}}$ whenever $X_1|X_2| \cdots| X_{k}\succ^b Y_1|Y_2| \cdots |Y_{l}$ with $Y_1=X_1X_2\cdots X_q$.
		\item If 
		\bea\label{coarsening-decreasing}
		d_{Z_1Z_2\cdots Z_s; Z_{s+1}}\geq d_{Z_{i_1}Z_{i_2}\cdots Z_{i_t}; Z_{s+1}}
		\eea holds for 
		any $Z_1|Z_2|\cdots |Z_s\succ^a Z_{i_1}|Z_{i_2}|\cdots |Z_{i_t}$, then 
		\bea\label{coarsening-decreasing2}
		D_{X_1;X_2; \cdots; X_{k}}\geq D_{Y_1;Y_2; \cdots; Y_{l}}
		\eea 
		for any $X_1|X_2| \cdots| X_{k}\succ^a Y_1|Y_2| \cdots |Y_{l}$, where $Z_1|Z_2|\cdots |Z_s$, $Z_{i_1}|Z_{i_2}|\cdots |Z_{i_t}$., $X_1|X_2| \cdots| X_{k}$ and $Y_1|Y_2| \cdots |Y_{l}$ are partitions of $A_1A_2\cdots A_n$ or subsystem of $A_1A_2\cdots A_n$.
	\end{itemize}
That is, if Eq.~(\ref{coarsening-decreasing}) holds true, then $D_{A_1;A_2;\cdots;A_n}$ is a complete multiparty quantum discord.
\end{theorem}

\begin{proof}
	For any $n$, analogous to that of tripartite and four-partite cases, all these coarsening relations can be derived 
	by taking the optimal measurement for the former one [e.g., in relation $D_{A;B;C;D}\geq D_{A;BCD}$, we assume that $D_{A;B;C;D}$ is obtained by the optimal measurement, i.e., $D_{A;B;C;D}= -S_{ABCD}+S_A+S_{AB}'-S_A'
	+S_{ABC}''-S_{AB}''+S_{ABCD}'''-S_{ABC}'''=(S_{ABCD}'''-S_{ABC}''' -S_{ABCD}+S_{ABC} )
	+(S_{ABC}''-S_{AB}'' -S_{ABC}+S_{AB} )
	+(S_{AB}'-S_{A}'-S_{AB} +S_{A} )$], and then by using either the fact that the mutual information is decreased under local operation (which is equivalent to $S'_{XY}-S'_{X}-S_{XY}+S_{X}\geq0$), or such a fact togeter with the assumption~(\ref{coarsening-decreasing}).
	The former four cases can be derived without the assumption~(\ref{coarsening-decreasing}) since all these cases can be
	reduced to $S'_{XY}-S'_{X}-S_{XY}+S_{X}\geq0$.
	The last case can be easily processed under the assumption~(\ref{coarsening-decreasing}).	
\end{proof}

Using similar arguments as that of Proposition~\ref{tripartite-complete-monogamy} and Proposition~\ref{monogamy-4}, we can conclude the following theorem easily.

\begin{theorem}\label{th2}
	The multipartite discord $D_{A_1;A_2;\cdots;A_n}$ is completely monogamous if it is complete, or equivalently, if it is monotonic under coarsening of subsystem(s),. 	
\end{theorem}

The former four monotonicity relations in Theorem~\ref{th1} are true automatically. 
So, in Theorem~\ref{th2}, we only need the other
monotonicity conditions as Eq.~(\ref{coarsening-decreasing2}). 
(Here, we should note that Eq.~(\ref{coarsening-decreasing2}) contains the second and the third monotonicity relations in Theorem~\ref{th1}
as special cases.)
We thus obtain the following proposition.

\begin{pro}\label{pro8}
	The multipartite discord $D_{A_1;A_2;\cdots;A_n}$ is completely monogamous provided that
	the condition~(\ref{coarsening-decreasing2}) is valid.
\end{pro}

Although $D_{A_1;A_2;\cdots;A_n}$ is not continuous (since $D_{A_1;A_2}$ is not continuous~\cite{Xu2010nc,Mazzola2010prl}), $n>2$,
we still can, taking the tripartite case for example, get the following monogamy relation:
If $D_{A;B;C}$ is monotonic under discord of subsystems, then for any given $\rho\in\mathcal{S}^{ABC}$
there exists
$0<\alpha<\infty$ such that
\bea\label{power}
D^{\alpha}_{A;B;C}\geq  D^{\alpha}_{A;B}
+ D^{\alpha}_{A;C}+ D^{\alpha}_{B;C}.
\eea
That is, $\alpha$ is dependent not only on $\dim \mH^{ABC}$ but also on
the given state.

\section{The coarser relation of global quantum discord}

In Ref.~\cite{Braga2012pra}, the following monogamy bound is proved provided that the bipartite discord does not
increase under loss of subsystems, i.e., $D_{A_1\cdots A_k:A_{k+1}}\geq D_{A_1:A_{k+1}}$, $2\leq k<n$:
\bea\label{monogamy-of-GQD}
{D}_{ A_1 : \cdots :A_n } \geq \sum_{k=1}^{n-1} {D}_{ A_1 : A_{k+1} }.
\eea
However, the assumption $D_{A_1\cdots A_k:A_{k+1}}\geq D_{A_1:A_{k+1}}$ is not valid in general. For example, consider a three qubit state
$\rho_{ABC} = (1/2) ( |000 \rangle \langle 000| + |+11 \rangle \langle +11 |)$,
where $|+\rangle = (|0\rangle + |1\rangle)/\sqrt{2}$,
it is evident that $D_{AB:C}=0< D_{A:C}$.
That is, $D_{A_1\cdots A_k:A_{k+1}}$ and $D_{A_1:A_{k+1}}$ are uncomparable since
the associated measurements are not compatible.

For any state $\rho\in\mS^{ABC}$, we let
$D_{A:B:C}(\rho)=S_{ABC}'-S_A'-S_B'-S_C'-S_{ABC}+S_A+S_B+S_C$ for some local von Neumann 
measurement $\Phi(\rho)=\rho'$.
It follows that
\beax
D_{A:B:C}-D_{A:BC}&\geq&( S_{ABC}'-S_A'-S_B'-S_C'-S_{ABC}+S_A+S_B+S_C) \\
&&-(S_{ABC}'-S_A'-S_{BC}'-S_{ABC}+S_A+S_{BC})\\
&=&(S_B+S_C-S_{BC})-(S_B'+S_C'-S_{BC}')\geq0
\eeax
since mutual information is decreased under local von Neumann measurement.
With the notations as in Section~\ref{4}, we let $X_1|X_2| \cdots |X_{k}$ and $Y_1|Y_2| \cdots |Y_{l}$ be two arbitrarily given partitions of $A_1A_2\cdots A_m$ or subsystem of $A_1A_2\cdots A_m$, and assume that $X_1|X_2| \cdots| X_{k}\succ^b Y_1|Y_2| \cdots |Y_{l}$. Then we can conclude
$D_{X_1:X_2:\cdots:X_{k}}\geq D_{Y_1:Y_2:\cdots:Y_{l}}$.

We now begin to discuss the coarsening relation of GQD as a complement to the results in Ref.~\cite{Braga2012pra}.
For any state $\rho^{XY}\in\mS^{XY}$, we denote $S(\rho^{XY}\|\rho^X\otimes\rho^Y)$ by
$S_{XY\|X\otimes Y}$ for brevity, where $S(\rho\|\sigma)=\tr (\rho\log_2\rho-\rho\log_2\sigma)$ is the relative entropy.
For any state $\rho\in\mS^{A_1A_2\cdots A_n}$, we write state after measurement $\Phi$ as in Eq.~(\ref{measurement-GQD})
as $\rho'$, write $S({\rho'^{XY}\|\rho'^{X}\otimes\rho'^{Y}})$ as $S'_{XY\|X\otimes Y}$, and write $I(\rho)-I(\rho')$ as $d^\Phi_{A_1:A_2:\cdots :A_n}$. Then, it is easy to argue that, for any $k<n$,
\bea\label{trade-off-GQD1}
&&d^\Phi_{A_1:A_2:\cdots :A_n}-d^\Phi_{A_{i_1}:A_{i_2}:\cdots :A_{i_k}}\nonumber\\
&=&S_{A_1A_2\cdots A_n\|A_{i_1}A_{i_2}\cdots A_{i_k} \otimes{A_{j_1}}\otimes{A_{j_2}}\otimes\cdots\otimes {A_{j_{n-k}}}}\nonumber\\
&&-S'_{A_1A_2\cdots A_n\|A_{i_1}A_{i_2}\cdots A_{i_k} \otimes{A_{j_1}}\otimes{A_{j_2}}\otimes\cdots\otimes {A_{j_{n-k}}}}.
\eea
where $j_s\neq i_t$, $1\leq t\leq k$, $1\leq j\leq n-k$.
With these notations in mind, we can easily conclude the following coarsening relations.

\begin{pro}\label{GQD-1}
	(i) Using the notations as Eq.~(\ref{trade-off-GQD1}), if $d^\Phi_{A_1:A_2:\cdots :A_n}\geq d^\Phi_{A_{i_1}:A_{i_2}:\cdots :A_{i_k}}$ for any $k<n$, then $D_{A_1:A_2:\cdots :A_n}$ is monotonic under discord of subsystems.
	(ii) $D_{X_1:X_2:\cdots:X_{k}}\geq D_{Y_1:Y_2:\cdots:Y_{l}}$ whenever $X_1|X_2| \cdots| X_{k}\succ^b Y_1|Y_2| \cdots |Y_{l}$. 
	Namely, $D_{A_1:A_2:\cdots :A_n}$ is complete whenever $d^\Phi_{A_1:A_2:\cdots :A_n}\geq d^\Phi_{A_{i_1}:A_{i_2}:\cdots :A_{i_k}}$ for any $k<n$.
\end{pro}

Next, we consider the complete monogamy of $D_{A_1:A_2:\cdots :A_n}$.
Take a three qubit state
$\rho_{ABC} = (1/2) ( |000 \rangle \langle 000| + |1+1 \rangle \langle 1+1 |)$ as in Ref.~\cite{Braga2012pra},
where $|+\rangle = (|0\rangle + |1\rangle)/\sqrt{2}$.
For this state, it is shown in Ref.~\cite{Braga2012pra} that
${D}_{ A : B : C } ={D}_{ A : B }\approx 0.204$ and
${D}_{ A : C }= 0$.
It is easy to see that ${D}_{B : C }={D}_{A : B}$.
That is, ${D}_{ A : B : C } ={D}_{ A : B }$ but ${D}_{ B : C }\neq 0$.
Namely, GQD violates the dis-correlated condition.
We thus obtain the following theorem.

\begin{theorem}
	${D}_{ A_1 : \cdots :A_n }$ is not completely monogamous.
\end{theorem}

That is, even though GQD is complete, it is not completely monogamous.
Comparing with MQD, MQD $D_{A_1;A_2;\cdots;A_n}$
is better than GQD $D_{A_1:A_2:\cdots :A_n}$ in such a sense.
Of course that, $D_{A_1:A_2:\cdots :A_n}$ has some merit, e.g., it  is symmetric but $D_{A_1;A_2;\cdots;A_n}$ is not symmetric (see Fig.~\ref{tab:table1} for more detail).

\section{Conclusions and discussions}

\begin{table}
	\caption{\label{tab:table1} Comparison between MQD, GQD and complete multipartite entanglement measure for tripartite case.
	CM is the abbreviation of `complete monogamous'.}	
	\begin{center}
		{\small
		\begin{tabular}{cccccc}\hline\hline
		Measure& \textbf{U1} &\textbf{U2} & \textbf{U3}&\textbf{U4}&CM   \\ \hline
$D_{A_1;A_2;A_3}$&$\checkmark$&$\times$& Partially, conditionally$^{a}$&$\checkmark$&Conditionally$^{b}$ \\ 
$D_{A_1:A_2:A_3}$&$\checkmark$& $\checkmark$ &Conditionally$^{c}$&$\checkmark$& $\times$\\
			 $E^{(3)}_f$&$\checkmark$&$\checkmark$ &$\checkmark$&$\checkmark$& $\checkmark$\\
			\hline\hline
		\end{tabular}}
	\end{center}
{\scriptsize
	\vspace{-2mm}
\hspace{17mm} $^{a}$ {See Theorem~\ref{th1}.}

\hspace{17mm} $^{b}$ {See Theorem~\ref{th2}.}

\vspace{-2mm}
\hspace{17mm} $^{c}$ {See Proposition~\ref{GQD-1}.}
}
\end{table}

In this paper we have 
discussed the monogamy relation of multipartite quantum discord and global quantum discord
in detail. Different from the monogamy scenario of quantum discord discussed in the previous literatures, we defined complete multiparty quantum discord and by which we put forward the complete monogamy framework for both multipartite quantum discord and global quantum discord.
In such a framework, to characterize the distribution of quantum discord,
the first issue is to check whether the quantity is monotonic under coarsening of subsystems and the other issue it to investigate the dis-correlate condition. The dis-correlate condition of quantum discord is the counterpart to the disentanglement condition for entanglement.
With the assumption of monotonicity under coarsening of subsystems, we proved that the multipartite quantum discord is completely monogamous. We presented counterexamples which implies that global quantum discord is not completely monogamous.
This fact also supports that the multipartite quantum discord in Ref.~\cite{Radhakrishman2020prl} is an excellent
generalization of quantum discord. Going further,
we conjecture that the assumptions in Theorem~\ref{th1}, Theorem~\ref{th2}, Proposition~\ref{pro8} and Proposition~\ref{GQD-1} are true but it seems difficult to prove and remains further investigation in the future.
In addition, our framework can be used for any multiparty nonlocal quantum correlation whenever the associated monogamy relation is concerned.


\ack{
This work is supported by the National Natural Science Foundation of
China under Grant No.~11971277, the Program for
the Outstanding Innovative Teams of Higher Learning Institutions of
Shanxi, the Scientific Innovation Foundation of the Higher
Education Institutions of Shanxi Province under Grants No.~2019KJ034, No.~2019L0742, and No.~2020L0471, the Natural Science Foundation of
Shanxi Province under Grant No.~201801D121016, and the Science Technology Plan Project of Datong City, China under Grants No.~2018151 and No.~2020153.
}	




\section*{References}

\begin{thebibliography} {99}
	
	
	\bibitem{ollivier2001quantum} Ollivier H and Zurek W H 2001
	Quantum discord: a measure of the quantumness of correlations
	\textit{Phys. Rev. Lett.} \textbf{88} 017901
	
	\bibitem{henderson2001classical} Henderson L and Vedral V 2001
	Classical, quantum and total correlations
	\textit{J. Phys. A: Math. Theor.} \textbf{34} 6899
	
	\bibitem{Datta2008prl} Datta A, Shaji A, and Caves C M 2008
	Quantum discord and the power of one qubit
	\textit{Phys. Rev. Lett.} \textbf{100} 050502
	
	\bibitem{DiVincenzo2004prl} Datta A and 
	Gharibian S 2009 Signatures of nonclassicality in mixed-state quantum computation
	\textit{Phys. Rev. A} \textbf{79} 042325
		
	\bibitem{Sarandy2009pra} Sarandy M S 2009 Classical correlation and quantum discord in critical systems.
	\textit{Phys. Rev. A} \textbf{80} 022108
	
	\bibitem{Xu2010nc} Xu J S, Xu X Y, Li C F \textit{et al} 2010
	Experimental investigation of classical and quantum correlations under decoherence
	\textit{Nat. Commun.} \textbf{1} 7 
	
	\bibitem{Rulli2011pra} Rulli C C and Sarandy M S 2011 Global quantum discord in multipartite systems
	\textit{Phys. Rev. A} \textbf{84} 042109		
	
	\bibitem{Giorgi2011prl} Giorgi G L, Bellomo B, Galve F \textit{et al} 2011 Genuine quantum and classical correlations in multipartite Systems
	\textit{Phys. Rev. Lett.} \textbf{107} 190501
		
	\bibitem{Modi2012rmp} Modi K, Brodutch A, Cable H \textit{et al} 2012 The classical-quantum boundary for correlations: Discord and related measures
	\textit{Rev. Mod. Phys.} \textbf{84} 1655 
		
	\bibitem{Bera2017epp} Bera A, Das T, Sadhukhan D \textit{et al} 2018 Quantum discord and its allies: a review
	\textit{Rep. Prog. Phys.} \textbf{81} 024001
	
	\bibitem{Radhakrishman2020prl} Radhakrishnan C, Lauri{\`e}re M and Byrnes T 2020
	Multipartite generalization of quantum discord
	\textit{Phys. Rev. Lett.} \textbf{124} 110401
	
	\bibitem{Hu2018pr} Hu M L, Hu X, Wang J \textit{et al} 2018 Quantum coherence and geometric quantum discord
	Phys. Rep. \textbf{762} 1 
	
	\bibitem{Zhou2019pra} Zhou H, Yuan X, Ma X 2019
	Unification of quantum resources in distributed scenarios
	Phys. Rev. A 99 022326
				
	\bibitem{Coffman} Coffman V, Kundu J and  Wootters W K 2000
	Distributed entanglement
	\textit{Phys. Rev. A} \textbf{61} 052306
	
	\bibitem{streltsov2012are} Streltsov A, Adesso G, Piani M \textit{et al} 2012
	Are general quantum correlations monogamous?
	\textit{Phys. Rev. Lett.} \textbf{109} 050503
	
	\bibitem{Dhar} Dhar H S, Pal A K, Rakshit D \textit{et al} 2017
	Monogamy of quantum correlations-a review
	\textit{In Lectures on General Quantum Correlations and their Applications}(Cham, Switzerland: Springer)
	
	\bibitem{GG} Gour G and  Guo Y 2018
	Monogamy of entanglement without inequalities
	\textit{Quantum} \textbf{2} 81 
	
	\bibitem{GG2019} Guo Y and Gour G 2019
	Monogamy of the entanglement of formation
	\textit{Phys. Rev. A} \textbf{99} 042305 
	
	\bibitem{G2020} Guo Y and Zhang L 2020 
	Multipartite entanglement measure and complete monogamy relation
	\textit{Phys. Rev. A} \textbf{101} 032301 
	
	\bibitem{Gzy2020} Guo Y, Zhang L and Yuan H 2020
	Entanglement measures induced by fidelity-based distances
	\textit{Quant. Inf. Process.} \textbf{19} 282 
	
	\bibitem{Eltschka2019quantum} Eltschka C and Siewert J 2018 
	Distribution of entanglement and correlations in all finite dimensions
	\textit{Quantum} \textbf{2} 64 
	
	
	
	\bibitem{Deng} Deng X, Xiang Y, Tian C \textit{et al} 2017
	Demonstration of monogamy relations for Einstein-Podolsky-Rosen steering in Gaussian cluster state
	\textit{Phys. Rev. Lett.} \textbf{118} 230501 
	
	\bibitem{Braga2012pra} Braga H C, Rulli C C, Oliveira T R de \textit{et al} 2012
	Monogamy of quantum discord by multipartite correlations
	\textit{Phys. Rev. A} \textbf{86} 062106  
	
	\bibitem{Ren2013qic} Ren X J and Fan H 2013
	Non-monogamy of quantum discord and upper bounds for quantum correlation
	\textit{Quant. Inf. Comput.} \textbf{13}(5-6) 0469-0478	
	
	\bibitem{Prabhu2pra} Prabhu R, Pati A K, SenDe A \textit{et al} 2012
	Conditions for monogamy of quantum correlations: Greenberger-Horne-Zeilinger versus W states
	\textit{Phys. Rev. A} \textbf{85}
	040102(R)
	
	\bibitem{Giorgi2011pra} Giorgi G L 2011
	Monogamy properties of quantum and classical correlations
	\textit{Phys. Rev. A} \textbf{84} 054301
	
	
	
	
	
		
	
	
	
	
	
	
			
	
	
	
		
	
	
		
	\bibitem{Fanchini2013pra} Fanchini F F \textit{et al} 2013
	Why the Entanglement of Formation is not generally monogamic 
	\textit{Phys. Rev. A} \textbf{87} 032317
	
	\bibitem{Bai2013pra} Bai Y K \textit{et al} 2013
	Exploring multipartite quantum correlations with the square of quantum discord
	\textit{Phys. Rev. A} \textbf{88} 012123
	
	
	\bibitem{Groisman2005pra} Groisman B, Popescu S and Winter A 2005 
	Quantum, classical, and total amount of correlations in a quantum state
	\textit{Phys. Rev. A} \textbf{72} 032317 
	
	\bibitem{Horodecki2009} Horodecki R, Horodecki P, Horodecki M and Horodecki K 2009
	Quantum entanglement
	\textit{Rev. Mod. Phys.} \textbf{81} 865 
	
	\bibitem{Nielsen} Nielsen M A, Chuang I L 2000 \textit{Quantum Computatation and
	Quantum Information} (Cambridge: Cambridge University Press)
		
	\bibitem{Mazzola2010prl} Mazzola L, Piilo J and  Maniscalco S 2010 
	Sudden transition between classical and quantum decoherence
\textit{Phys. Rev. Lett.} \textbf{104} 200401  
	
	
	
\end{thebibliography}

\end{document}